\documentclass[conference]{IEEEtran}
\IEEEoverridecommandlockouts
\usepackage{cite}
\usepackage{amsmath,amssymb,amsfonts}
\usepackage{algorithmic}
\usepackage{graphicx}
\usepackage{textcomp}
\usepackage{xcolor}
\usepackage{subcaption}
\usepackage{float}
\newcommand{\commentout}[1]{}
\def\BibTeX{{\rm B\kern-.05em{\sc i\kern-.025em b}\kern-.08em
    T\kern-.1667em\lower.7ex\hbox{E}\kern-.125emX}}
\begin{document}

\title{Scaling Wideband Massive MIMO Radar via Beamspace Dimension Reduction \\
}

\author{\IEEEauthorblockN{
Oveys Delafrooz Noroozi\IEEEauthorrefmark{1}, Jiyoon Han\IEEEauthorrefmark{2}, Wei Tang\IEEEauthorrefmark{2}, Zhengya Zhang\IEEEauthorrefmark{2}}, Upamanyu Madhow\IEEEauthorrefmark{1}
\IEEEauthorblockA{
\IEEEauthorrefmark{1}University of California, Santa Barbara, CA, U.S.A., \IEEEauthorrefmark{2}University of Michigan, Ann Arbor, MI, U.S.A.}
\{oveys, madhow\}@ucsb.edu
\{hanjyoon, weitang, zhengya\}@umich.edu}

\maketitle

\begin{abstract}
We present an architecture for scaling digital beamforming for wideband massive MIMO radar.  Conventional spatial processing becomes computationally prohibitive as array size grows; for example, the computational complexity of MVDR beamforming scales as $O(N^3)$ for an $N$-element array.  In this paper, we show that energy concentration in beamspace provides the basis for drastic complexity reduction, with array scaling governed by the $O(N \log N)$ complexity of the spatial FFT used for beamspace transformation.  Specifically, we propose an architecture for windowed beamspace MVDR beamforming, parallelized across targets and subbands, and evaluate its efficacy for beamforming and interference suppression for government-supplied wideband radar data from the DARPA SOAP (Scalable On-Array Processing) program.  We demonstrate that our approach achieves detection performance comparable to full-dimensional benchmarks while significantly reducing computational and training overhead, and provide insight into tradeoffs between beamspace window size and FFT resolution in balancing complexity, detection accuracy, and interference suppression.

%Beamspace processing—implemented via a spatial FFT across uniform antenna arrays—has long served as an effective dimension-reduction technique in narrowband array signal processing, enabling efficient implementations of direction-of-arrival estimation, adaptive beamforming, and multiuser MIMO. However, extending beamspace techniques to wideband radar systems introduces new challenges due to the frequency-dependent array response, unlike the narrowband case where the response is effectively constant across the signal bandwidth. Addressing these challenges is critical for realizing scalable digital signal processing architectures for wideband massive MIMO radar, where conventional algorithms like MVDR beamforming become computationally prohibitive as array size grows. This paper presents a scalable architecture for applying windowed beamspace MVDR beamforming to wideband radar systems with large antenna arrays. We evaluate our proposed architecture on a government-supplied dataset from the DARPA SOAP (Scalable On-Array Processing) program, demonstrating that windowed beamspace MVDR achieves detection performance comparable to full-dimensional benchmarks while significantly reducing computational and training overhead. The results further highlight key trade-offs between beamspace window size and FFT resolution in balancing complexity, detection accuracy, and interference suppression.
\end{abstract}

\section{Introduction}
\label{Sec:Introduction}

In contrast to the scanning beams provided by the classical phased arrays employed in conventional radar systems, fully digital arrays offer the possibility of forming simultaneous beams for acquisition and/or tracking multiple targets, while utilizing some of the available spatial degrees of freedom for suppressing clutter and interference.  Advances in silicon realizations of both radio frequency (RF) and digital signal processing (DSP) imply that the hardware for realizing massive digital arrays (e.g., with 100s or 1000s of elements) for wideband massive MIMO radar is within reach. However, the computational complexity of conventional DSP algorithms presents a fundamental bottleneck to such scaling:  for example, classical narrowband minimum variance distortionless response (MVDR) beamforming incurs $O(N^3)$ complexity, where $N$ denotes the number of antennas.  The recently launched DARPA SOAP program asks the question: is it possible to attain a more attractive computational scaling of $O(N \log N)$ for wideband massive MIMO radar?  In this paper, we answer this question in the affirmative, proposing and evaluating on government furnished data (GFD) a modern, massively parallelized DSP architecture which exploits the classical concept of beamspace dimension reduction.

For regularly spaced antenna arrays, the array response for a signal arriving from a given direction is a complex exponential, so that its energy can be concentrated via a spatial FFT, also termed a {\it beamspace} transformation.  Beamspace techniques have been extensively explored across radar and array signal processing applications as a means to enhance robustness and reduce computational complexity. Applications span from through-the-wall radar imaging \cite{yoon2008high, yoon2011mvdr}, weather radar applications \cite{nai2016adaptive}, and moving target localization via beamspace MVDR \cite{jin2002beamspace, wang2008mvdr} to subarray-based approaches that mitigate array imperfections and mismatches such as sensor position errors \cite{doisy2010interference}. Early contributions in this area include partial adaptation schemes and beamspace implementations for MUSIC that leverage dimension reduction for improved performance \cite{zoltowski1990simultaneous}.  These ideas pave the way for the scalable wideband adaptive beamforming architecture proposed here. 

Our approach is also informed by recent work on beamspace dimension reduction for multiuser MIMO communication: energy concentration in beamspace allows the use of a small beamspace window for each user, large enough to capture most of the energy of the user of interest and to provide the necessary dimensions for suppressing ``nearby'' interference. Recent work \cite{abdelghany2019beamspace, abdelghany2020scalable} demonstrates that such processing achieves performance close to conventional full-dimensional ``antenna space'' methods, with \cite{cebeci2024scaling} further validating the approach using real-world experimental data. 

In this paper, we are interested in scaling MVDR beamforming for processing wideband radar signals to massive digital arrays.  The MVDR beamformer aims to minimize the output power of the array while maintaining a distortionless response in a desired signal direction.  The classical least squares implementation requires estimation and inversion of the sample covariance matrix, incurring $O(N^3)$ computational complexity.  Classical MVDR beamforming operates under the so-called narrowband assumption, in which the array response is well-modeled as invariant across the signal bandwidth.  Wideband signals can be channelized into subbands, with MVDR beamformers computed and applied separately for each subband.  Our proposed architecture is also based on such channelization, with drastic complexity reduction obtained by employing windowed beamspace MVDR in each subband.

\noindent
{\bf Approach and Contributions:}
We consider wideband radar returns corresponding to a sequence of chirps, decomposed into subbands. For each subband, a beamspace transformation is applied, followed by dimension reduction using fixed windowing.  Depending on implementation considerations, the order of channelization and beamspace transformation can be commuted, since they are both linear operations. Reduced-dimension MVDR beamforming is then performed independently for each subband, and the outputs are synthesized to recover the wideband signal before standard range-Doppler processing. These operations are performed in parallel for each target of interest.   

We evaluate the performance of windowed beamspace MVDR processing on simulated radar data provided as GFD.  We analyze the trade-offs between beamspace window size and FFT resolution, highlighting their impact on detection performance and interference suppression. The results demonstrate that our approach maintains effective interference suppression and detection performance while significantly reducing computational and training costs, showing the feasibility of digital beamforming and wideband radar processing using massive antenna arrays.

\section{System Model}
\label{Sec:System Model}
\subsection{Received Signal Model}\label{Sec:Signal}

We consider a wideband radar system employing a two-dimensional uniform planar array (UPA) with \( N_z \) elements along the vertical (z-axis) and \( N_x \) elements along the horizontal (x-axis), resulting in a total of \( N = N_z \times N_x \) antenna elements. The array elements are uniformly spaced with inter-element spacing \( d = \lambda/2 \), where \( \lambda \) is the wavelength corresponding to the design frequency \( f_d \).

The radar transmits linear frequency-modulated chirps and receives a combination of reflections from desired targets and direct transmissions from external interferers, such as communication base stations or other radar systems. The received data consists of discrete-time in-phase and quadrature (IQ) samples after downconversion.

Assuming a total of \(L\) subbands, the center frequency of subband \(\ell\), where \(\ell = -L/2 + 1, \ldots, L/2\), is given by
\begin{equation}
    f(\ell) = f_c + \left( \frac{\ell - 1/2}{L} \right) f_s,
    \label{Eq:SubChannelFreq}
\end{equation}
where \( f_c \) is the carrier frequency and \( f_s \) is the sampling rate.

\begin{figure}[t]
    \centering
    \includegraphics[width=1\linewidth, height=0.6\linewidth]{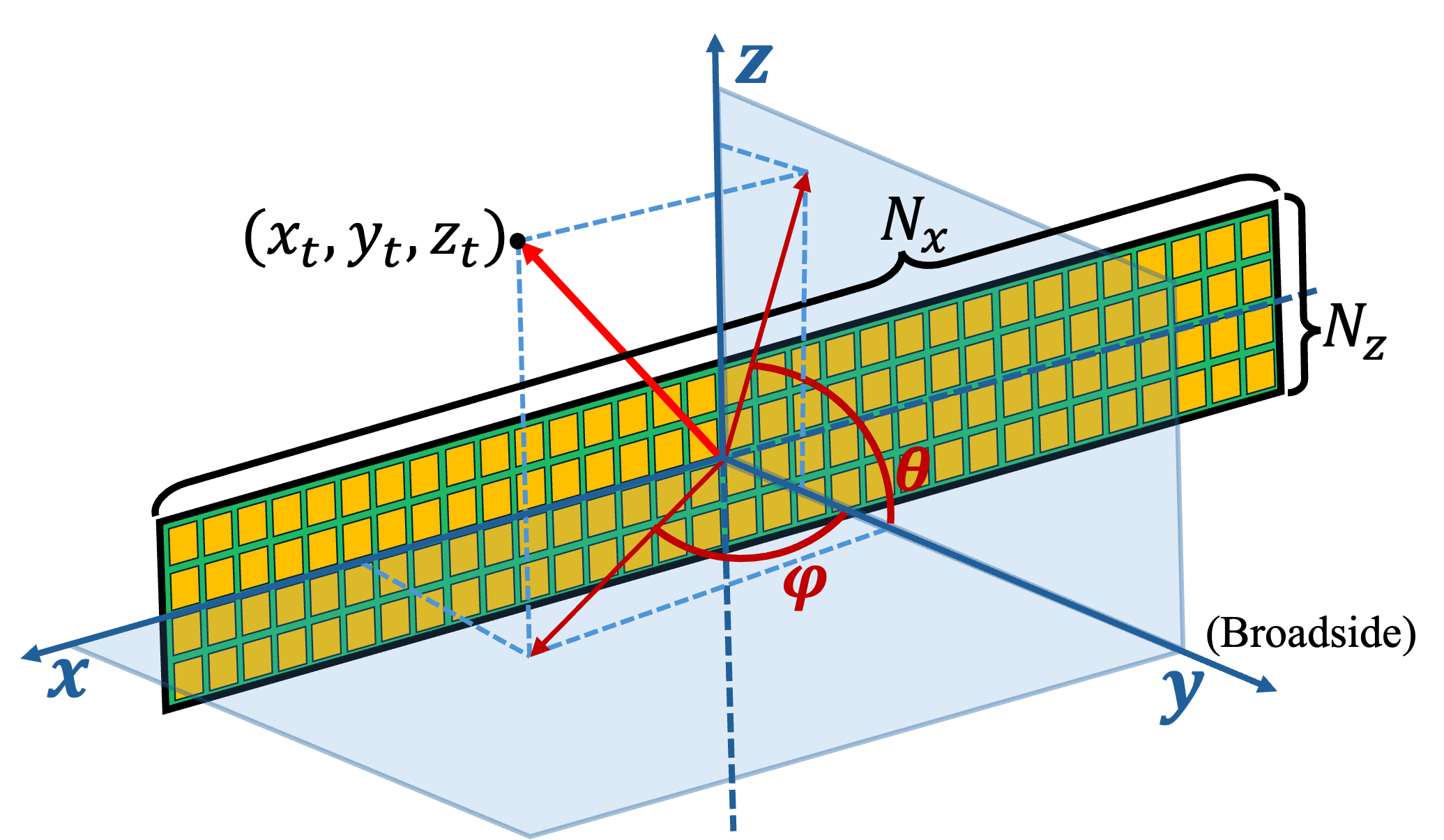}
      \caption{\small Antenna array geometry showing the broadside direction and definitions of elevation (\(\theta\)) and azimuth (\(\varphi\)) angles corresponding to the target location \((x_t, y_t, z_t)\).}
    \label{Fig:Geometry}
\end{figure}

As depicted in Fig. \ref{Fig:Geometry}, the broadside direction of the array is aligned with the y-axis, corresponding to \((\varphi, \theta) = (0^\circ, 0^\circ)\). Throughout this paper, both the azimuth angle \(\varphi\) and elevation angle \(\theta\) are defined with respect to the y-axis.
\begin{itemize}
    \item \(\varphi\) denotes the azimuth angle measured from the y-axis toward the x-axis,
    \item \(\theta\) denotes the elevation angle measured from the y-axis toward the z-axis.
\end{itemize}

The reference spatial frequencies, defined with respect to the design frequency \( f_d \), are given by
\begin{align}
    \Omega_{x,k}^{\text{ref}} &= \pi \cos\theta_k \sin\varphi_k,
    \label{Eq:OmegaX}\\
    \Omega_{z,k}^{\text{ref}} &= \pi \sin\theta_k.
    \label{Eq:OmegaZ}
\end{align}

At subband \(\ell\) centered at frequency \( f(\ell) \), the effective spatial frequencies are scaled as
\begin{align}
    \Omega_{x,k}^{(\ell)} &= \Omega_{x,k}^{\text{ref}} \times \frac{f(\ell)}{f_d},
    \label{Eq:SubOmegaX}\\
    \Omega_{z,k}^{(\ell)} &= \Omega_{z,k}^{\text{ref}} \times \frac{f(\ell)}{f_d}.
    \label{Eq:SubOmegaZ}
\end{align}

The vectorized steering vector for the \(k\)-th target at subband \(\ell\) is constructed as the Kronecker product:
\begin{equation}
    \mathbf{a}_{k}^{(\ell)} = \mathbf{a}_{x}(\Omega_{x,k}^{(\ell)}) \otimes \mathbf{a}_{z}(\Omega_{z,k}^{(\ell)}),
    \label{Eq:ArrayFactor}
\end{equation}
where
\begin{align}
    \mathbf{a}_{z}(\Omega_{z,k}^{(\ell)}) &= 
    \begin{bmatrix}
        1 & e^{j\Omega_{z,k}^{(\ell)}} & \cdots & e^{j\Omega_{z,k}^{(\ell)}(N_z-1)}
    \end{bmatrix}^T,
    \label{Eq:ArrayZ} \\
    \mathbf{a}_{x}(\Omega_{x,k}^{(\ell)}) &= 
    \begin{bmatrix}
        1 & e^{j\Omega_{x,k}^{(\ell)}} & \cdots & e^{j\Omega_{x,k}^{(\ell)}(N_x-1)}
    \end{bmatrix}^T.
    \label{Eq:ArrayX}
\end{align}

While the steering vector depends on the subband index \(\ell\), we omit it from the notation (i.e., use \(\mathbf{a}_k\)) throughout the paper for clarity, as the subsequent processing is identical across subbands.

Finally, the vectorized received signal at a given subband is denoted by \( \mathbf{y}[n] \in \mathbb{C}^{N} \), where \( N = N_z N_x \), and is modeled as:
\begin{equation}
    \mathbf{y}[n] = \sum_{k=1}^{K} \alpha_k \mathbf{a}_k p_k[n] + \mathbf{I}[n] + \mathbf{n}[n],
    \label{Eq:ReceivedSignal}
\end{equation}
where:
\begin{itemize}
    \item \(K\) is the total number of targets.
    \item \( \alpha_k \in \mathbb{C} \) is the complex channel coefficient for the \(k\)-th target,
    \item \( p_k[n] \in \mathbb{C} \) is the sampled baseband pulse of the \(k\)-th target,
    \item \( \mathbf{a}_k \in \mathbb{C}^{N} \) is the steering vector for the \(k\)-th target
    \item \( \mathbf{I}[n] \) models aggregate interference,
    \item \( \mathbf{n}[n] \sim \mathcal{CN}(\mathbf{0}, \sigma^2 \mathbf{I}) \) is spatially white Gaussian noise.
\end{itemize}

\subsection{Conventional MVDR Beamforming}\label{Sec:ConvMVDR}

The objective of MVDR beamforming is to suppress interference and noise while preserving the signal arriving from a desired spatial direction.

The theoretical covariance matrix is defined as
\begin{equation}
    \mathbf{R} = \mathbb{E} \left[ \mathbf{y}[n] \mathbf{y}^H[n] \right],
    \label{Eq:IdealCov}
\end{equation}
and its empirical estimate based on \( n_t \) snapshots is
\begin{equation}
    \widehat{\mathbf{R}} = \frac{1}{n_t} \sum_{n=1}^{n_t} \mathbf{y}[n] \mathbf{y}^H[n].
    \label{Eq:EmpiricalCov}
\end{equation}

In each subband, for a given target index \(k\), the MVDR correlator vector \( \mathbf{c}_k \) is computed by solving the following optimization problem:
\begin{equation}
    \begin{aligned}
    \min_{\mathbf{c}_k} \quad & \mathbf{c}_k^H \widehat{\mathbf{R}} \mathbf{c}_k \\
    \text{subject to} \quad & \mathbf{c}_k^H \mathbf{a}_k = 1.
    \end{aligned}
    \label{Eq:DefMVDR}
\end{equation}

The closed-form solution for the correlator vector is
\begin{equation}
    \mathbf{c}_k = \frac{ \widehat{\mathbf{R}}^{-1} \mathbf{a}_k }{ \mathbf{a}_k^H \widehat{\mathbf{R}}^{-1} \mathbf{a}_k }.
    \label{Eq:CorrMVDR}
\end{equation}

Applying the correlator \( \mathbf{c}_k \) to the received vector \( \mathbf{y}[n] \) (i.e. \(\mathbf{c}_k^H\mathbf{y}[n]\)) produces the beamformer output targeting direction \(k\).

\section{Proposed System}
\label{Sec:Proposed System}
\subsection{Overview}\label{Sec:Overview}

We propose a scalable windowed beamspace MVDR framework for wideband radar systems that reduces both computational complexity and training overhead while maintaining effective target detection. The key idea is to operate in a reduced-dimensional subspace that captures the dominant spatial components of the received signal, allowing efficient beamforming with minimal performance degradation.

\begin{figure}[t]
    \begin{subfigure}{0.5\textwidth}
        \centering
        \includegraphics[width=\linewidth, height=0.5\linewidth]{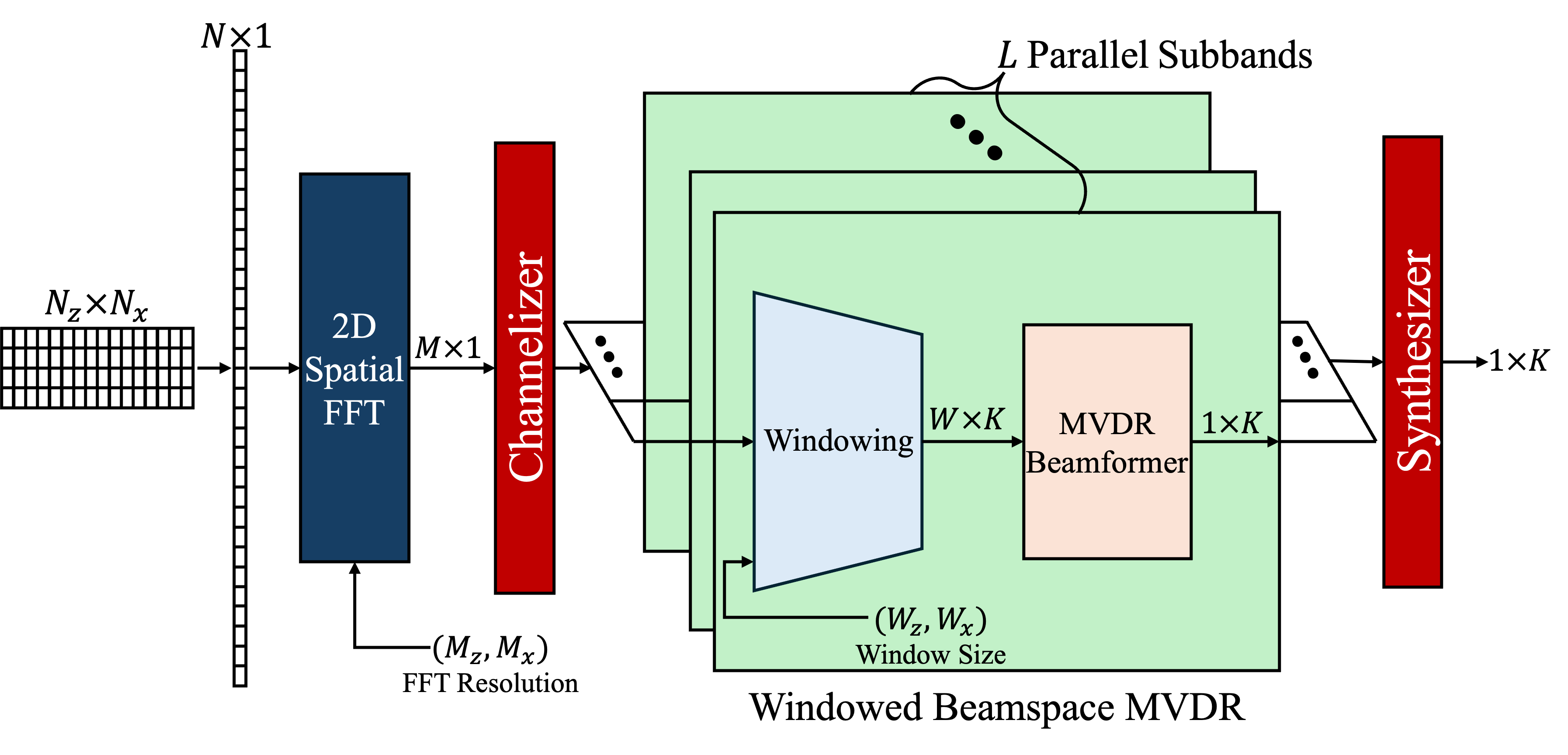}
        \caption{\small{System overview}}
        \label{Fig:SystemModel}
    \end{subfigure}
    \begin{subfigure}{0.5\textwidth}
        \centering
        \includegraphics[width=\linewidth, height=0.4\linewidth]{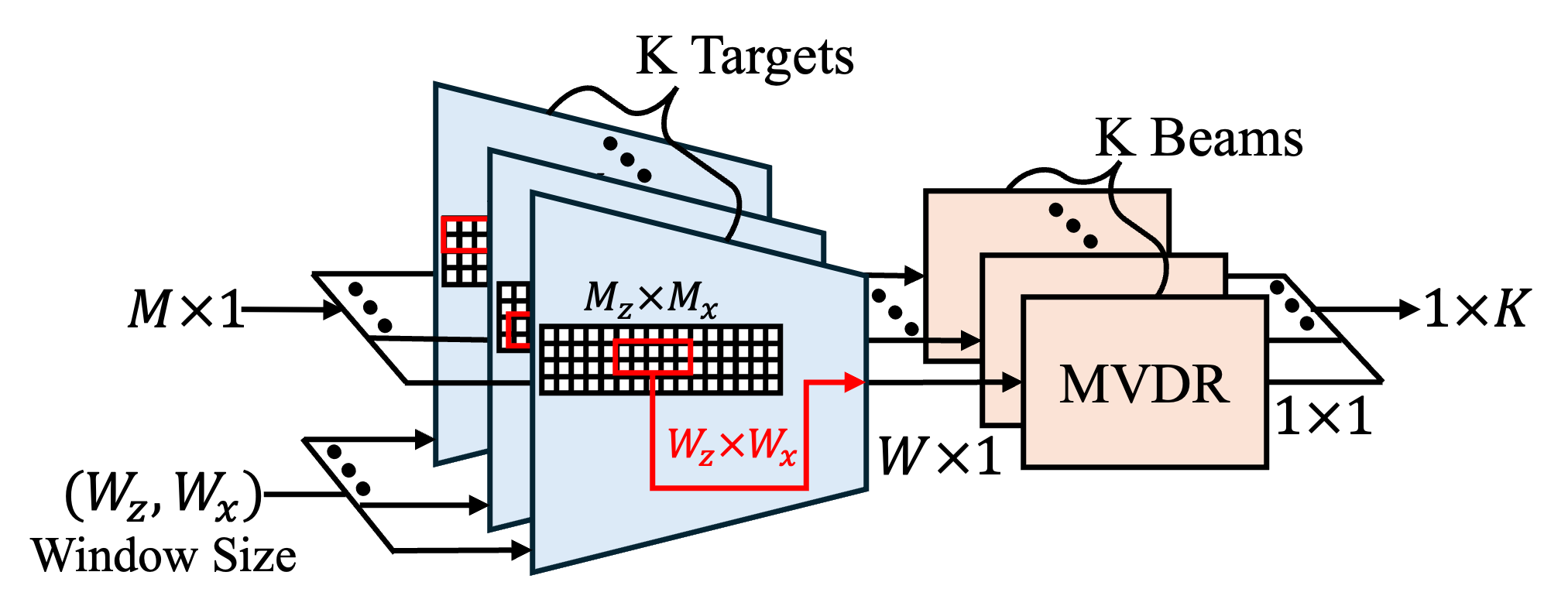}
        \caption{\small{Processing per subband}}
        \label{Fig:SysPerSub}
    \end{subfigure}    
    \caption{\small (a) Overall system pipeline including 2D spatial FFT, channelization, windowed beamspace MVDR, and signal synthesis.\\
    (b) Per-subband (narrowband) processing of windowed beamspace MVDR module, showing spatial window extraction of size \((W_z \times W_x)\) from a beamspace input of size \((M_z \times M_x)\) for each target, followed by MVDR beamforming.}
\end{figure}

Figures ~\ref{Fig:SystemModel} and \ref{Fig:SysPerSub} illustrate the overall processing pipeline and processing per subchannel before range-Doppler processing, respectively. The framework begins with applying a spatial 2D FFT—where the number of FFT points can be configured (i.e., with optional zero-padding)—to project the received signal into beamspace, enabling finer angular resolution through spatial oversampling. Then we partition the wideband received signal into multiple narrowband subbands using an FFT-based channelizer. To reduce the dimensionality, a fixed window selects a subset of beamspace components that captures most of the signal energy of each target in each subband. The resulting lower-dimensional signal is used as input to a subband-specific MVDR beamformer that performs interference suppression and spatial filtering in beamspace. The outputs of the beamformers across subbands are synthesized using an inverse FFT to recover the wideband signal prior to range-Doppler processing. This modular architecture enables frequency-dependent spatial filtering and scales efficiently with the number of antennas and subbands.

The following subsections provide details of the beamspace projection, windowing structure, and reduced-dimension MVDR formulation.

\subsection{Beamspace Transformation and Windowing}\label{Sec:BeamWindow}

To enable scalable wideband processing, we project the received signal at each subband into beamspace using a two-dimensional spatial Fourier transform. In general, we define the transform with a configurable number of FFT points in each dimension to allow for optional zero-padding, which we later evaluate as a design parameter. This operation is modeled using a pair of truncated DFT matrices that define the spatial transform.

We define truncated DFT matrices \( \mathbf{D}_v \in \mathbb{C}^{M_z \times N_z} \) and \( \mathbf{D}_h \in \mathbb{C}^{N_x \times M_x} \), which project the antenna space signal onto a denser spatial frequency grid of size \( M_z \times M_x \), where \( M_z \geq N_z \) and \( M_x \geq N_x \). The entries of these matrices are given by:

\begin{align}
[\mathbf{D}_v]_{m,n} &= \frac{1}{\sqrt{M_z}} e^{-j \frac{2\pi}{M_z} (m-1)(n-1)}, \nonumber \\
&\quad m = 1,\dots, M_z,\;\; n = 1,\dots, N_z,
\label{Eq:VerDFT}\\
[\mathbf{D}_h]_{n,m} &= \frac{1}{\sqrt{M_x}} e^{-j \frac{2\pi}{M_x} (m-1)(n-1)}, \nonumber \\
&\quad m = 1,\dots, M_x,\;\; n = 1,\dots, N_x.
\label{Eq:HorDFT}
\end{align}

The full 2D beamspace transform is then written as a Kronecker product:
\begin{equation}
\mathbf{D} = \mathbf{D}_h^T \otimes \mathbf{D}_v \in \mathbb{C}^{M \times N}, \quad M = M_z M_x.
\label{Eq:OverallDFT}
\end{equation}
Applying this transformation yields the beamspace signal:
\begin{equation}
\widehat{\mathbf{y}}[n] = \mathbf{D} \, \mathbf{y}[n] \in \mathbb{C}^{M}.
\label{Eq:BeamTransform}
\end{equation}

To reduce processing dimensionality, we apply a fixed window in beamspace. Let \( \mathbf{S}_{v,k} \in \mathbb{R}^{W_z \times M_z} \) and \( \mathbf{S}_{h,k} \in \mathbb{R}^{W_x \times M_x} \) denote the vertical and horizontal selector (windowing) matrices that extract a rectangular region of size \( W = W_z \times W_x \) centered on the FFT bin corresponding to target \(k\).

The full beamspace windowing matrix is then constructed as:
\begin{equation}
\mathbf{S}_k = \mathbf{S}_{h,k}^T \otimes \mathbf{S}_{v,k} \in \mathbb{R}^{W \times M}, \quad W = W_z W_x.
\label{Eq:OverallWindow}
\end{equation}

Applying this selector matrix to the beamspace signal yields the reduced-dimensional windowed beamspace signal:
\begin{equation}
\tilde{\mathbf{y}}_k[n] = \mathbf{S}_k \, \widehat{\mathbf{y}}[n] \in \mathbb{C}^{W}.
\label{Eq:ApplyWindow}
\end{equation}

\subsection{Reduced-Dimension MVDR}\label{Sec:BeamMVDR}

Given the dimension-reduced signal \( \tilde{\mathbf{y}}_k[n] \in \mathbb{C}^W \), we compute the beamspace MVDR weights using the same formulation as in Section~\ref{Sec:System Model}. Replacing \( \mathbf{y}[n] \) in Eq. \ref{Eq:OmegaX} with \( \tilde{\mathbf{y}}_k[n] \) yields the windowed beamspace covariance matrix \( \tilde{\mathbf{R}}_k \in \mathbb{C}^{W \times W} \), which captures the spatial structure of the reduced beamspace signal. Using this matrix in Eq. \ref{Eq:CorrMVDR}, we obtain the reduced-dimension MVDR correlator:

\begin{equation}
\tilde{\mathbf{c}}_k = \frac{ \tilde{\mathbf{R}}_k^{-1} \, \tilde{\mathbf{a}}_k }{ \tilde{\mathbf{a}}_k^H \tilde{\mathbf{R}}_k^{-1} \tilde{\mathbf{a}}_k } \in \mathbb{C}^{W},
\label{Eq:BeamCorrMVDR}
\end{equation}

where \( \tilde{\mathbf{a}}_k = \mathbf{S}_k \, \mathbf{D} \, \mathbf{a}_k \) is the windowed beamspace steering vector corresponding to target direction \(k\).

To interpret the beamspace MVDR output in the full antenna domain, we define the antenna-space equivalent correlator by lifting the windowed beamspace MVDR weights back through the adjoint of the transform:

\begin{equation}
\widehat{\mathbf{c}}_k = \mathbf{D}^H \, \mathbf{S}_k^T \, \tilde{\mathbf{c}}_k \in \mathbb{C}^{N}.
\label{Eq:AntennaEqCorr}
\end{equation}

This lifted weight vector enables direct comparison with full-dimensional antenna-space beamformers and helps interpret the beam pattern or correlator structure in the original domain. Additionally, in setups where the weight vector is obtained during a separate training phase, the lifted correlator can be directly applied during inference, eliminating the need for per-sample spatial FFT and windowing.

\section{performance evaluation}
\label{Sec:performance evaluation}
\subsection{Dataset and Evaluation Scenarios} \label{Sec:Dataset}
We evaluate the proposed approach using the GFD wideband radar dataset collected from a \(4 \times 32\) antenna array mounted on an aircraft flying at an altitude of 7~km. The aircraft travels eastward (along the \(x\)-axis) at a speed of 90~m/s, with the array oriented to face north (along the \(y\)-axis). The dataset includes 20 moving airborne targets, along with combinations of ground-based and sea-based interferers, such as communication and radar systems.

The evaluation spans five scenarios, labeled A through E, with the number of interferers increasing from Scenario A to Scenario E. Each scenario is provided in two geometric configurations: an \textit{Easy} mode (indicated by index 1), where targets maintain a larger angular separation from the ground—measured relative to a point located 18~km in front of the array along the \(y\)-axis—with elevation angles ranging from \(25^\circ\) to \(50^\circ\); and a \textit{Difficult} mode (indicated by index 2), where this separation is reduced to \(12.5^\circ\). For each scenario, a corresponding dataset in the \textit{Non-Interferer} mode is also available, containing only the targets with no interfering sources present.

\begin{figure}[t]
    \centering
    \includegraphics[width=\linewidth, height = 0.9\linewidth]{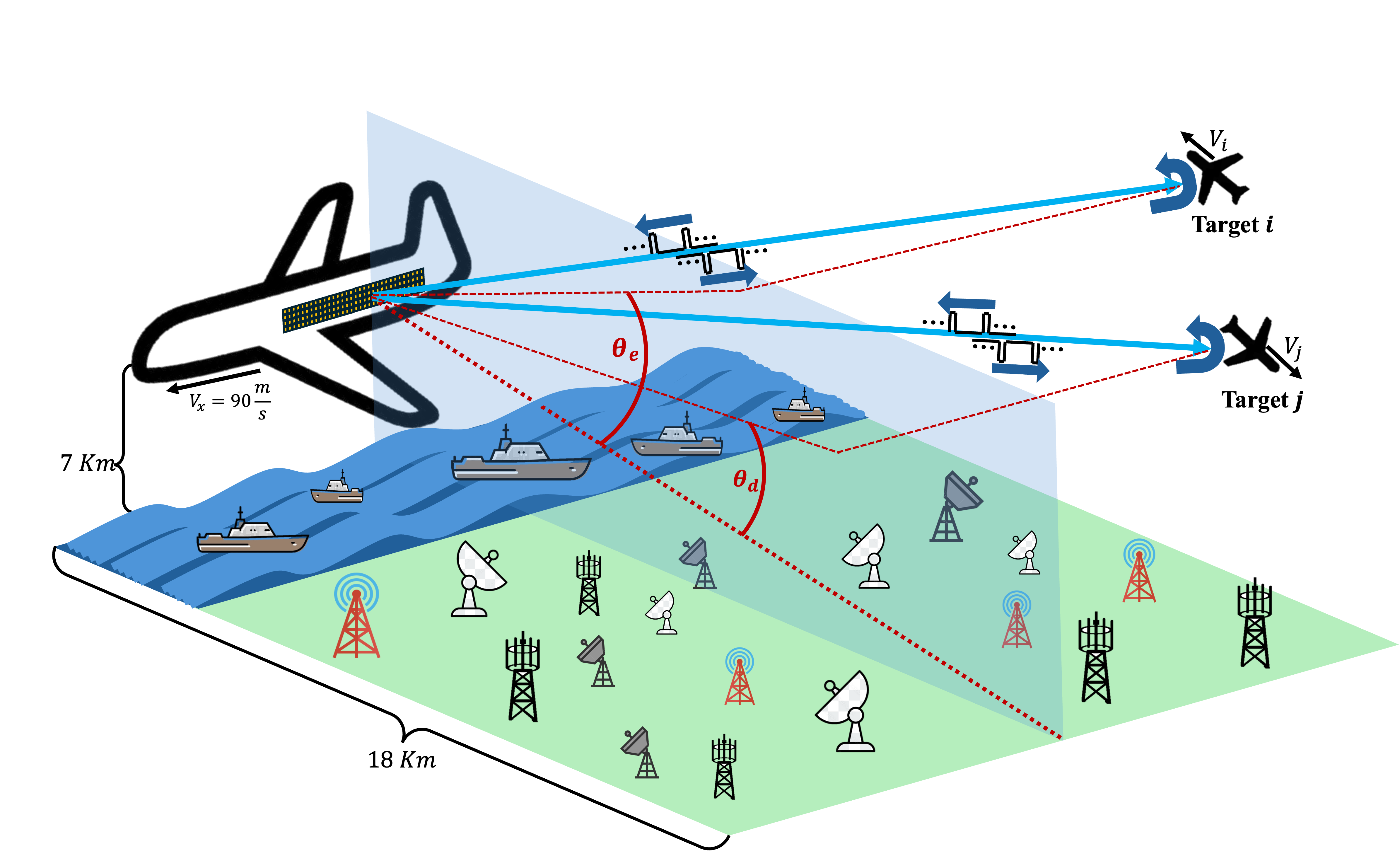}
    \caption{\small Simulated environment where an array-equipped platform observes multiple airborne targets, as well as ground-based and sea-based interferers. The elevation separations \(25^\circ \leq \theta_e \leq 50^\circ\) and \(\theta_d \geq 12.5^\circ\) correspond to relatively easy and difficult beamforming cases, respectively. Targets \(i\) and \(j\) exemplify easy and difficult scenarios.}
    \label{Fig:Environement}
\end{figure}

Figure~\ref{Fig:Environement} illustrates the simulated environment. Target \(i\) appears in an Easy-mode configuration, while target \(j\) represents a Difficult-mode case. The figure also shows representative ground- and sea-based interferers placed across the scene.

\subsection{Experimental Setup} \label{Sec:Setup}

We evaluate the proposed method using the GFD wideband radar dataset described in Section~\ref{Sec:Dataset}, which captures returns from a \(4 \times 32\) uniform planar array mounted on an airborne platform. The received signal is divided into \(L = 128\) narrowband subbands via FFT-based channelization, and each subband is processed independently using the proposed windowed beamspace MVDR framework.

For target detection, we apply a 1D CFAR processor across all velocity bins in the range-Doppler map. A constant false alarm threshold of 10~dB above the estimated noise floor is used to declare detection. Detection accuracy is evaluated in terms of range and velocity error, and we study how these results vary with window size and FFT resolution in beamspace.

\subsection{Numerical Results} \label{Sec:NumResults}

We evaluate performance in two contrasting conditions: the easiest scenario (\(A_1\)) and the most challenging scenario (\(E_2\)). Our analysis examines the impact of window size and FFT resolution (zero-padding) on detection accuracy across both range and velocity. Detection errors are quantized to the underlying range and velocity grid resolutions determined by the processing setup. 

\begin{figure}[t]
    \centering
    \begin{subfigure}{0.48\textwidth}
        \centering
        \includegraphics[width=\linewidth, height=0.6\linewidth]{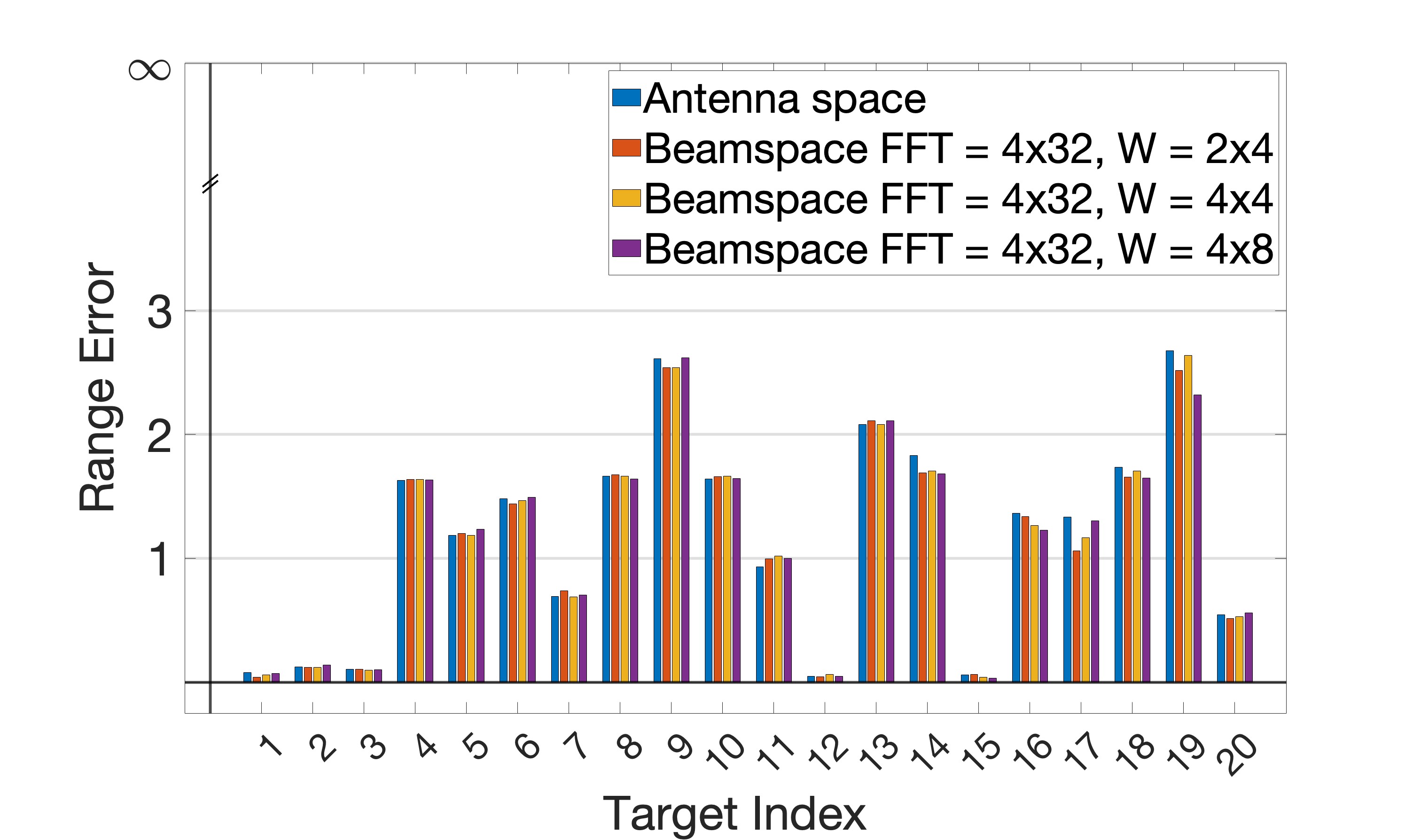}
        \caption{\footnotesize}
        \label{Fig:A1_Rng_W}
    \end{subfigure}
    \begin{subfigure}{0.48\textwidth}
        \centering
        \includegraphics[width=\linewidth, height=0.6\linewidth]{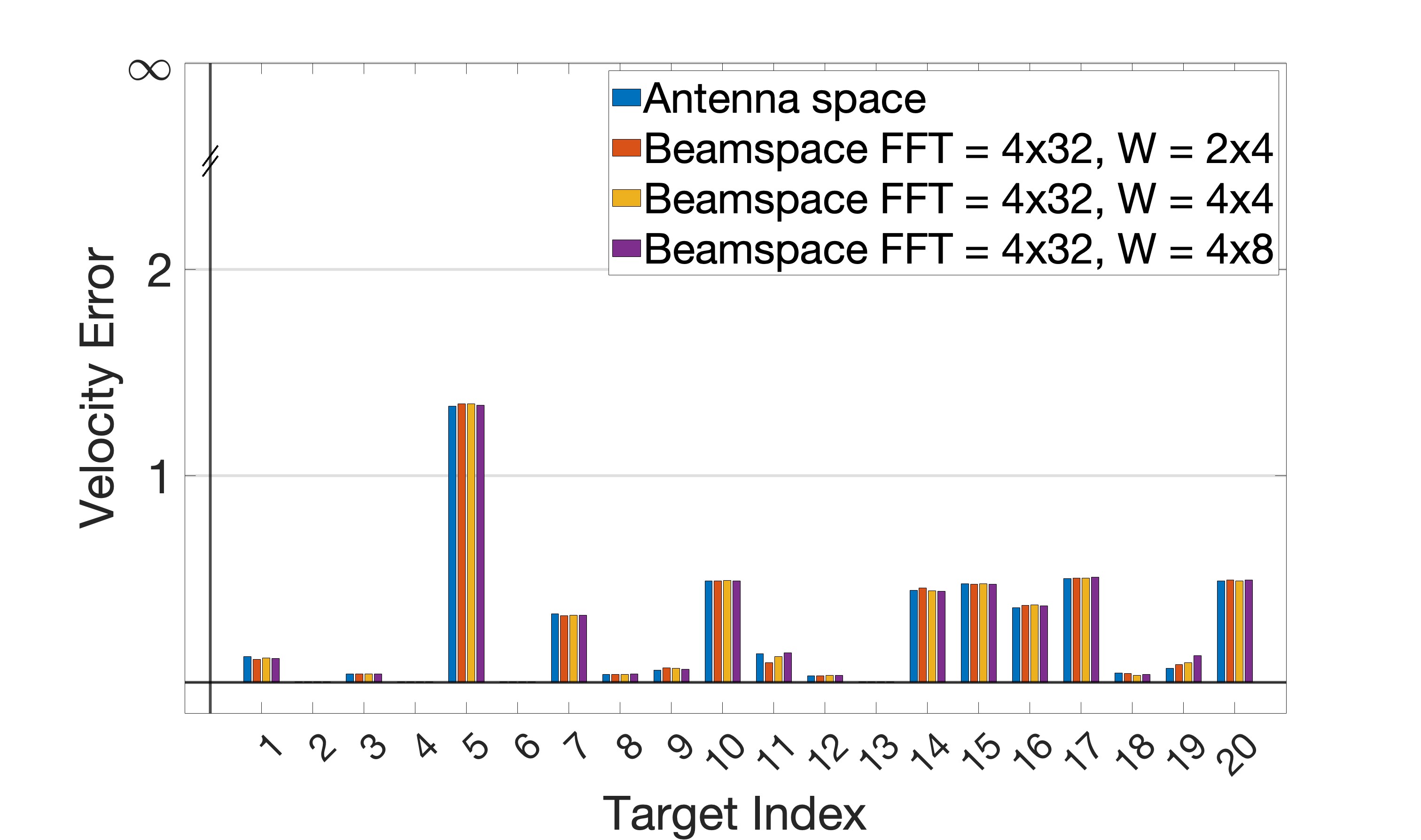}
        \caption{\footnotesize}
        \label{Fig:A1_Vel_W}
    \end{subfigure}
    \caption{\small Effect of window size on detection results for Scenario \(\text{A}_1\).}
    \label{Fig:A1_W}
\end{figure}

\begin{figure}[H]
    \centering
    % Second row of subfigures
    \begin{subfigure}{0.48\textwidth}
        \centering
        \includegraphics[width=\linewidth, height=0.6\linewidth]{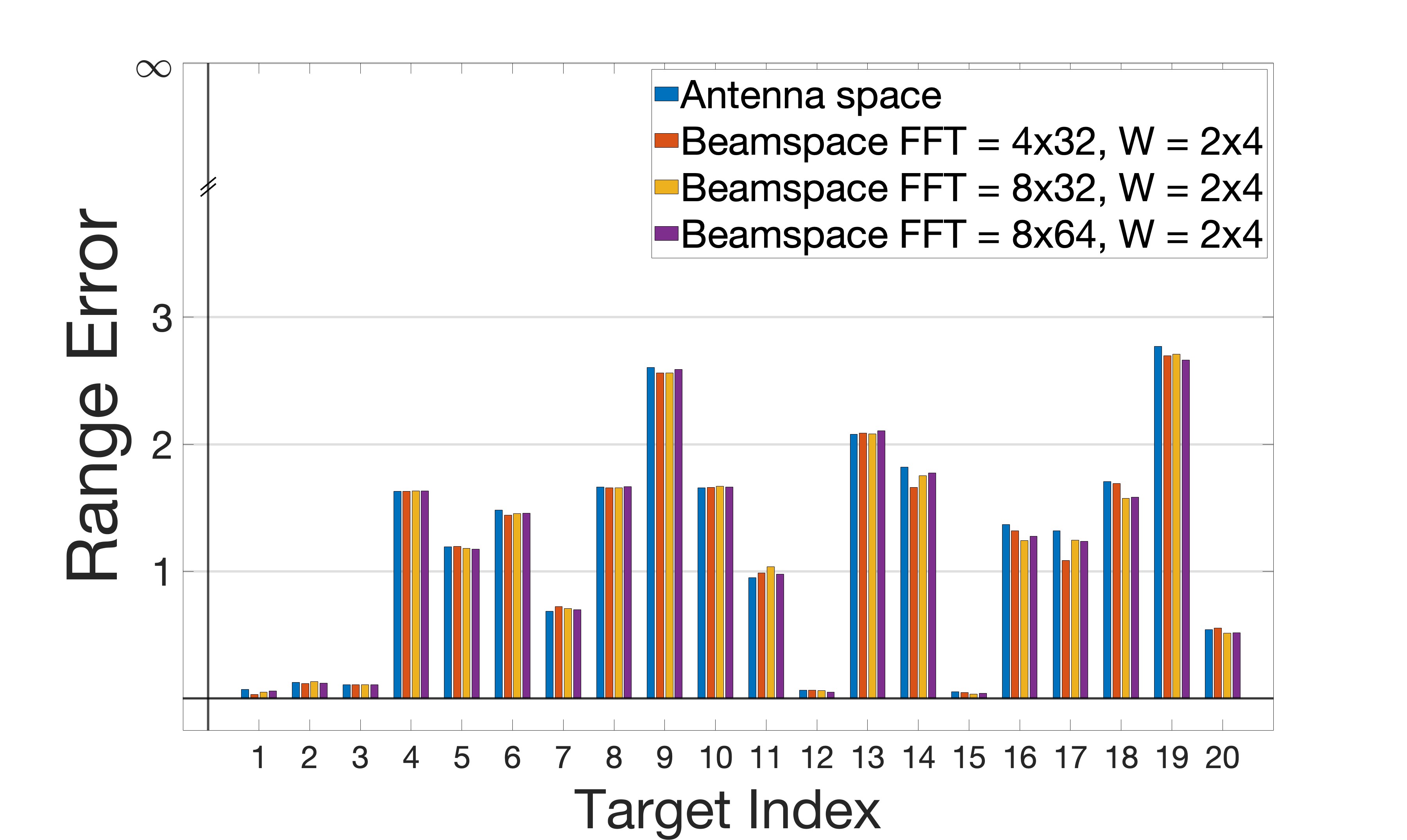}
        \caption{\footnotesize}
        \label{Fig:A1_Rng_F}
    \end{subfigure}
    \begin{subfigure}{0.48\textwidth}
        \centering
        \includegraphics[width=\linewidth, height=0.6\linewidth]{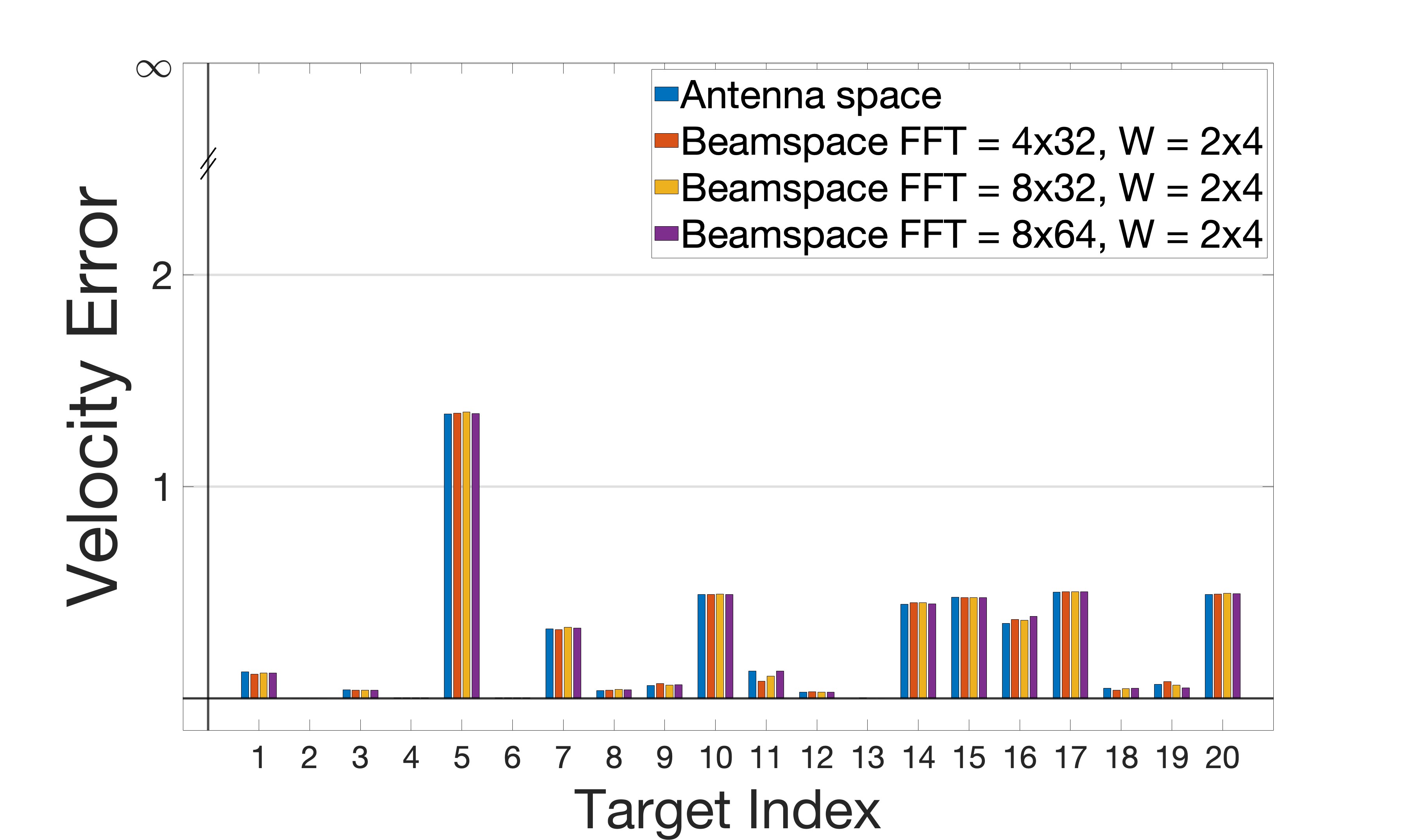}
        \caption{\footnotesize}
        \label{Fig:A1_Vel_F}
    \end{subfigure}
    \caption{\small Effect of zero-padding on detection results for Scenario \(\text{A}_1\).}
    \label{Fig:A1_F}
\end{figure}

Figures~\ref{Fig:A1_W} and~\ref{Fig:A1_F} show results for Scenario \(A_1\). Despite using a small beamspace window of \(2 \times 4\), the proposed method achieves detection accuracy comparable to or better than full-dimensional MVDR. In this low-interference regime, beamspace energy is well concentrated, and a small number of FFT bins captures the desired signal while inherently suppressing most interferers due to angular separation. As a result, increasing the window size provides no significant benefit and may even introduce additional interference leakage. Zero-padding also has negligible effect in this scenario, as beam alignment is already sufficient with basic FFT resolution.

\begin{figure}[t]
    \centering
    \begin{subfigure}{0.48\textwidth}
        \centering
        \includegraphics[width=\linewidth, height=0.6\linewidth]{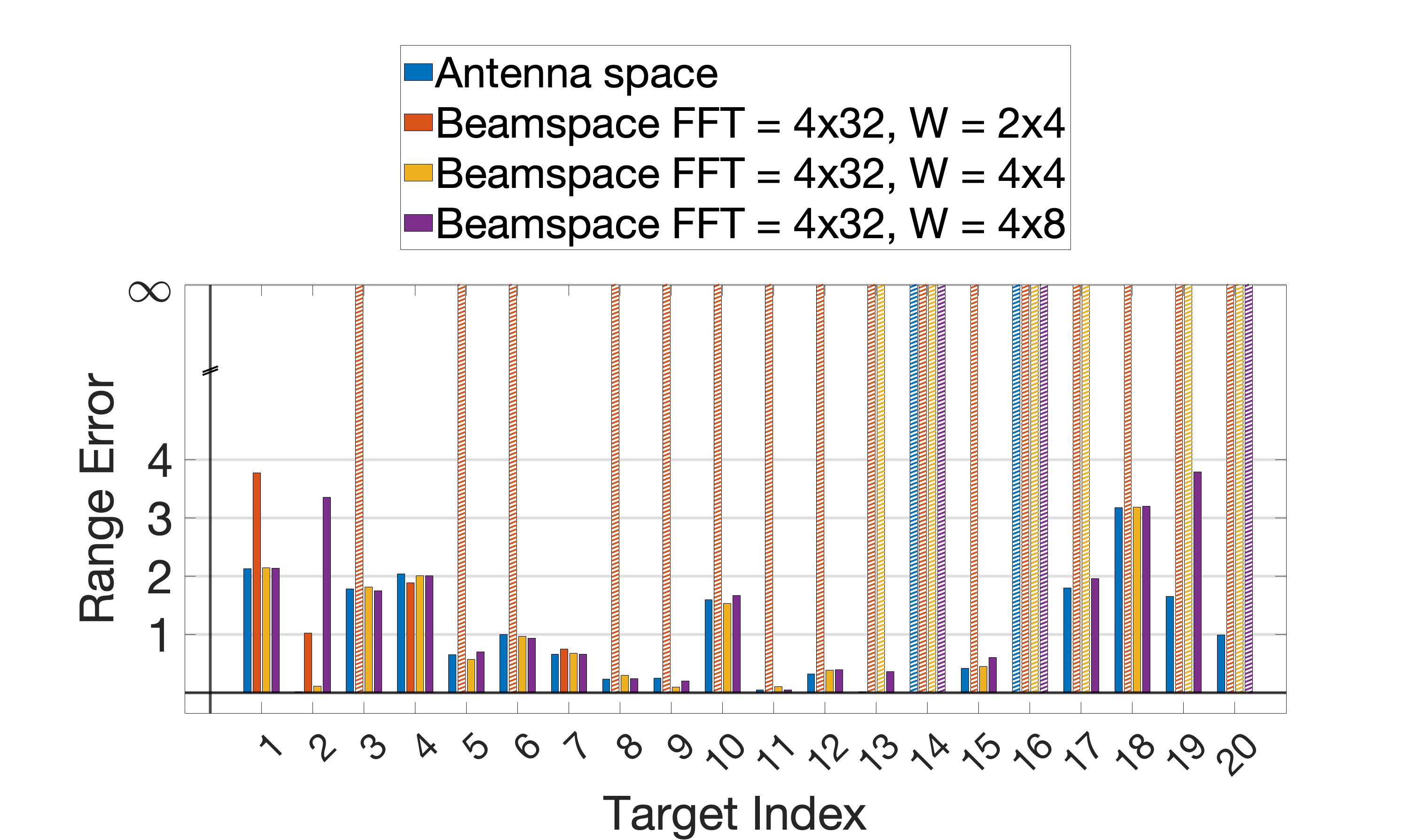}
        \caption{\footnotesize}
        \label{Fig:E2_Rng_W}
    \end{subfigure}
    \begin{subfigure}{0.48\textwidth}
        \centering
        \includegraphics[width=\linewidth, height=0.6\linewidth]{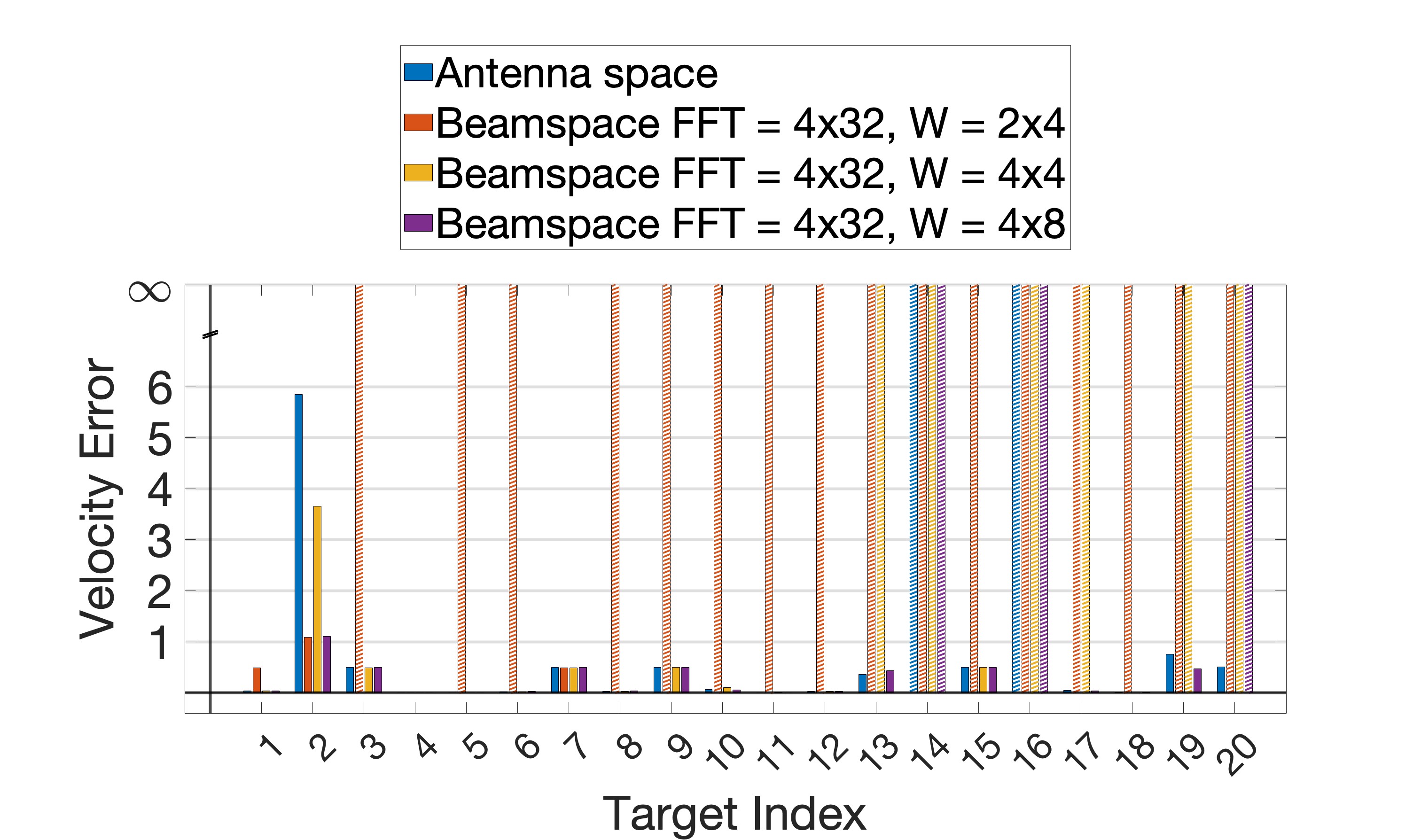}
        \caption{\footnotesize}
        \label{Fig:E2_Vel_W}
    \end{subfigure}
    \caption{\small Effect of window size on detection results for Scenario \(\text{E}_2\).}
    \label{Fig:E2_W}
\end{figure}

\begin{figure}[t]
    \centering
    % Second row of subfigures
    \begin{subfigure}{0.48\textwidth}
        \centering
        \includegraphics[width=\linewidth, height=0.6\linewidth]{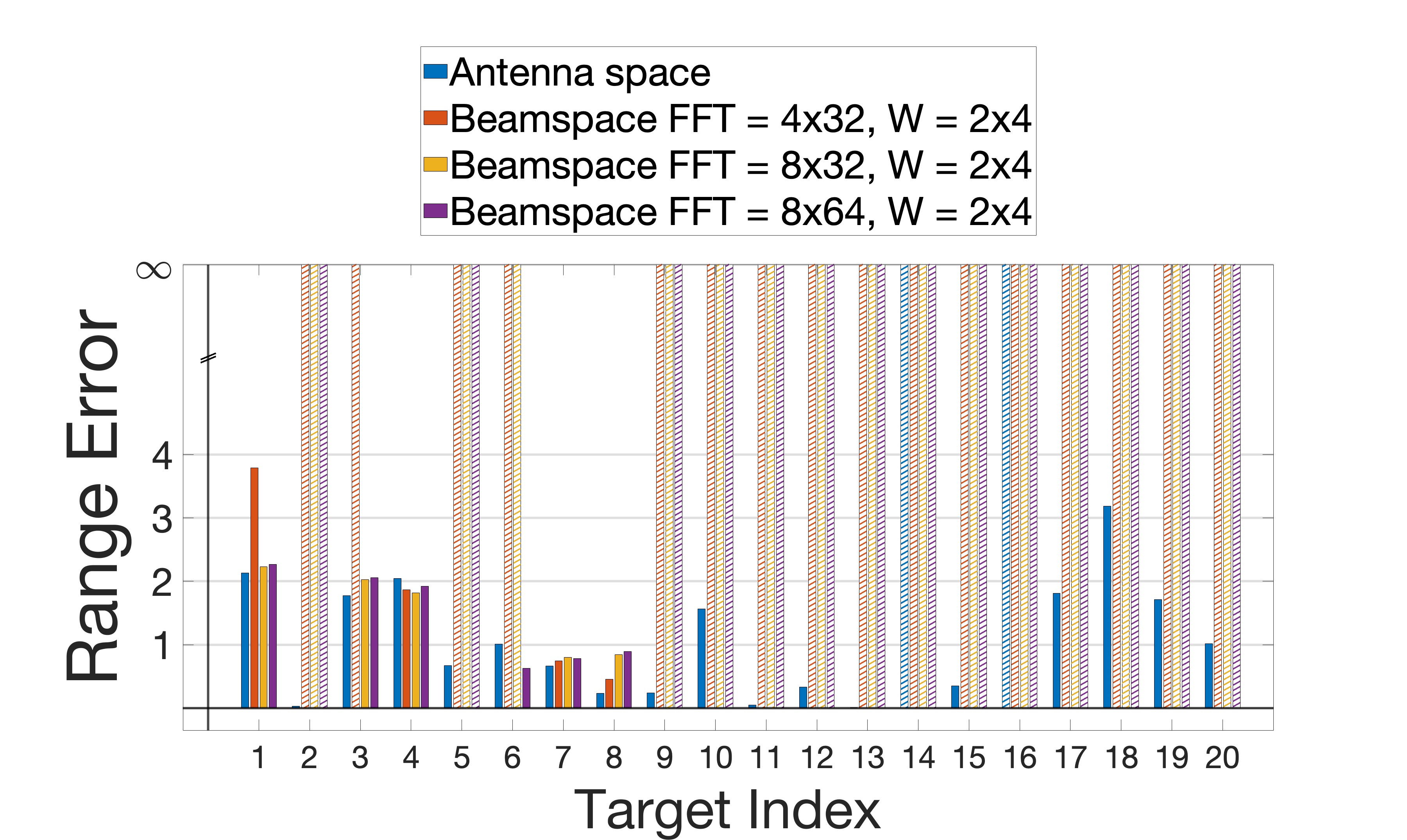}
        \caption{\footnotesize}
        \label{Fig:E2_Rng_F}
    \end{subfigure}
    \begin{subfigure}{0.48\textwidth}
        \centering
        \includegraphics[width=\linewidth, height=0.6\linewidth]{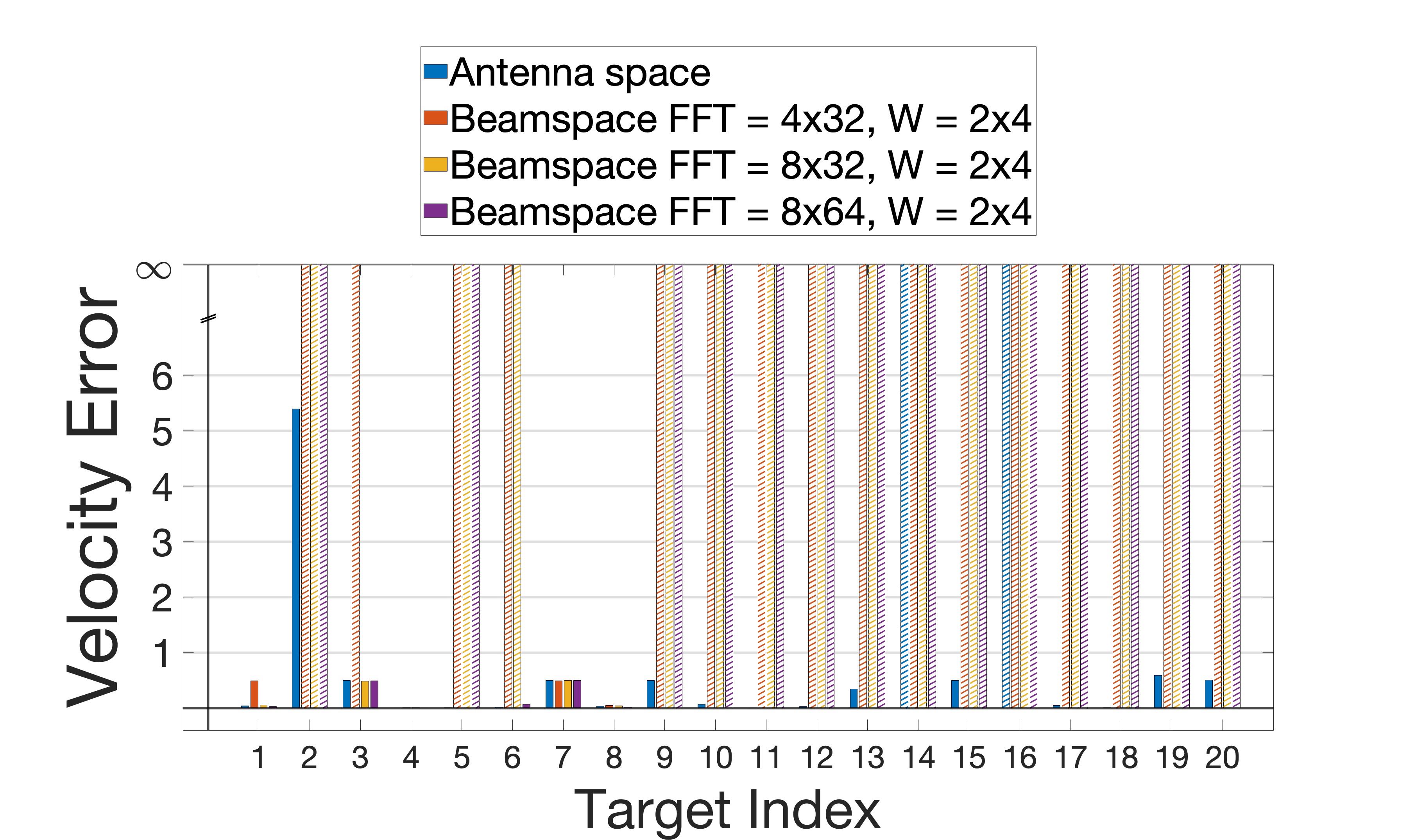}
        \caption{\footnotesize}
        \label{Fig:E2_Vel_F}
    \end{subfigure}
    \caption{\small Effect of zero-padding on detection results for Scenario \(\text{E}_2\).}
    \label{Fig:E2_F}
\end{figure}

In contrast, Figures~\ref{Fig:E2_W} and~\ref{Fig:E2_F} depict results for Scenario \(E_2\), where numerous strong interferers are present and the beamspace is densely occupied. In these figures, missed detections are indicated by hatched bars and assigned a symbolic infinite error to distinguish them from successfully detected targets. Under this high loading factor, the separation in the beam domain deteriorates, and a small window is insufficient to isolate and suppress interference. As shown in Figure~\ref{Fig:E2_W}, increasing the window sizes to \(4 \times 4\) or \(4 \times 8\)—still significantly smaller than full antenna space—substantially improves detection accuracy and roughly matches the number of detected targets achieved by full-dimensional MVDR.

Zero-padding also contributes in difficult scenarios by improving beam resolution and angular separation. For example, in Scenario \(E_2\), targets 3 and 6 are only detected when using an FFT size of \(8 \times 64\), whereas they are missed with lower FFT resolution. Nevertheless, this benefit is secondary compared to the impact of increased window size, which plays the dominant role in matching full-dimensional MVDR performance in both range and velocity estimates.

As an additional evaluation, we compare the beamforming patterns produced by antenna-space MVDR and windowed beamspace MVDR. The beamforming pattern for a given MVDR weight vector is defined using cosine similarity between the correlator and the steering vector:
\begin{equation}
    B_k(\varphi, \theta) = \frac{|\langle \mathbf{c}_k, \mathbf{a}(\varphi, \theta) \rangle|}{\|\mathbf{c}_k\|_2 \, \|\mathbf{a}(\varphi, \theta)\|_2}, \label{Cosine Sim}
\end{equation}
where \( \mathbf{c}_k \) is the MVDR weight vector for target \(k\), and \( \mathbf{a}(\varphi, \theta) \) is the steering vector corresponding to azimuth and elevation angles \( (\varphi, \theta) \). Equation~\ref{Eq:AntennaEqCorr} enables this comparison by expressing the beamspace correlator in the antenna domain. By substituting \(\widehat{\mathbf{c}}_k\) into Equation~\ref{Cosine Sim}, we obtain the beamforming pattern corresponding to the windowed beamspace MVDR.


\begin{figure}[t]
    \centering
    % Second row of subfigures
    \begin{subfigure}{0.48\textwidth}
        \centering
        \includegraphics[width=\linewidth, height=0.6\linewidth]{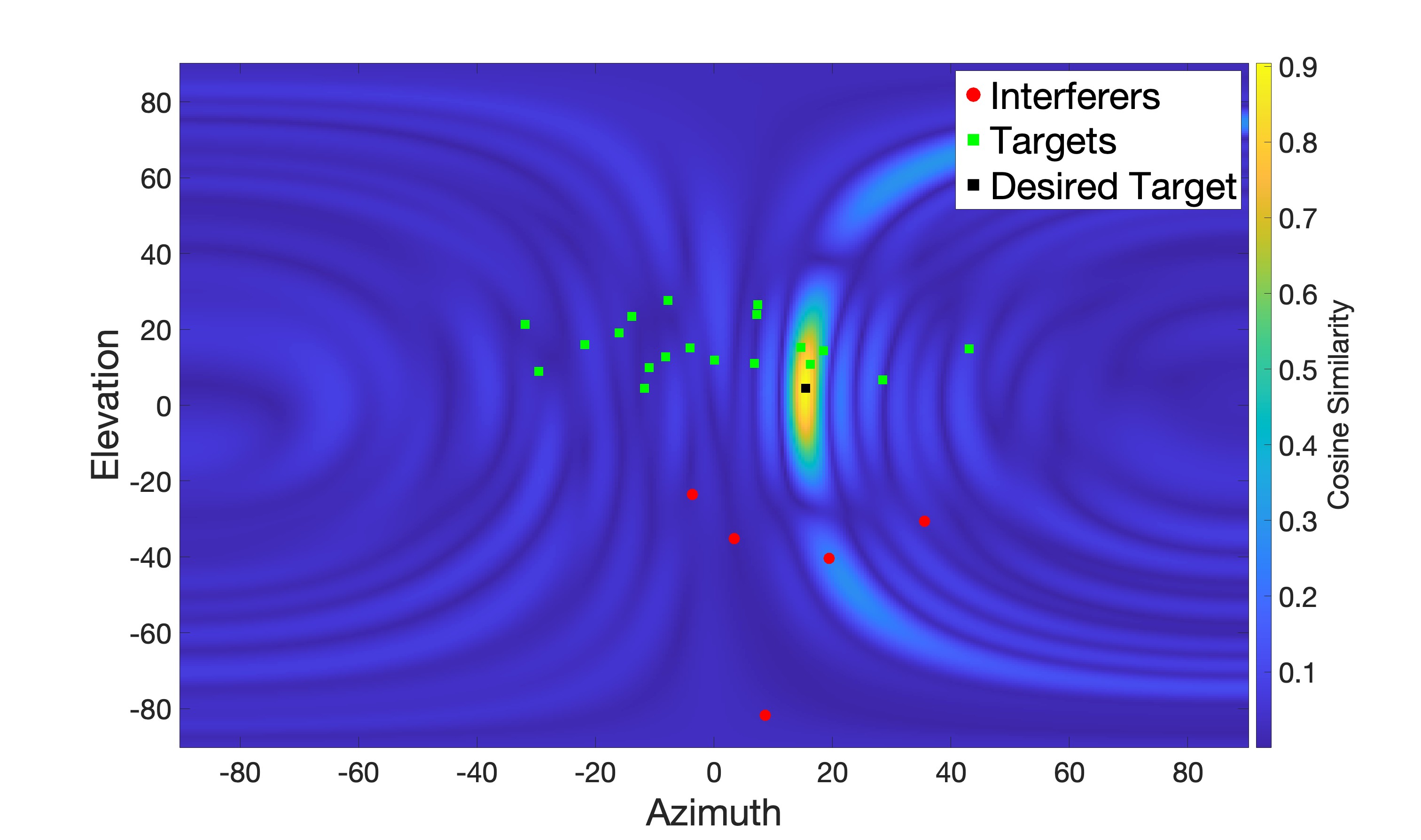}
        \caption{\footnotesize{Antenna space}}
        \label{Fig:BP_A1_A}
    \end{subfigure}
    \begin{subfigure}{0.48\textwidth}
        \centering
        \includegraphics[width=\linewidth, height=0.6\linewidth]{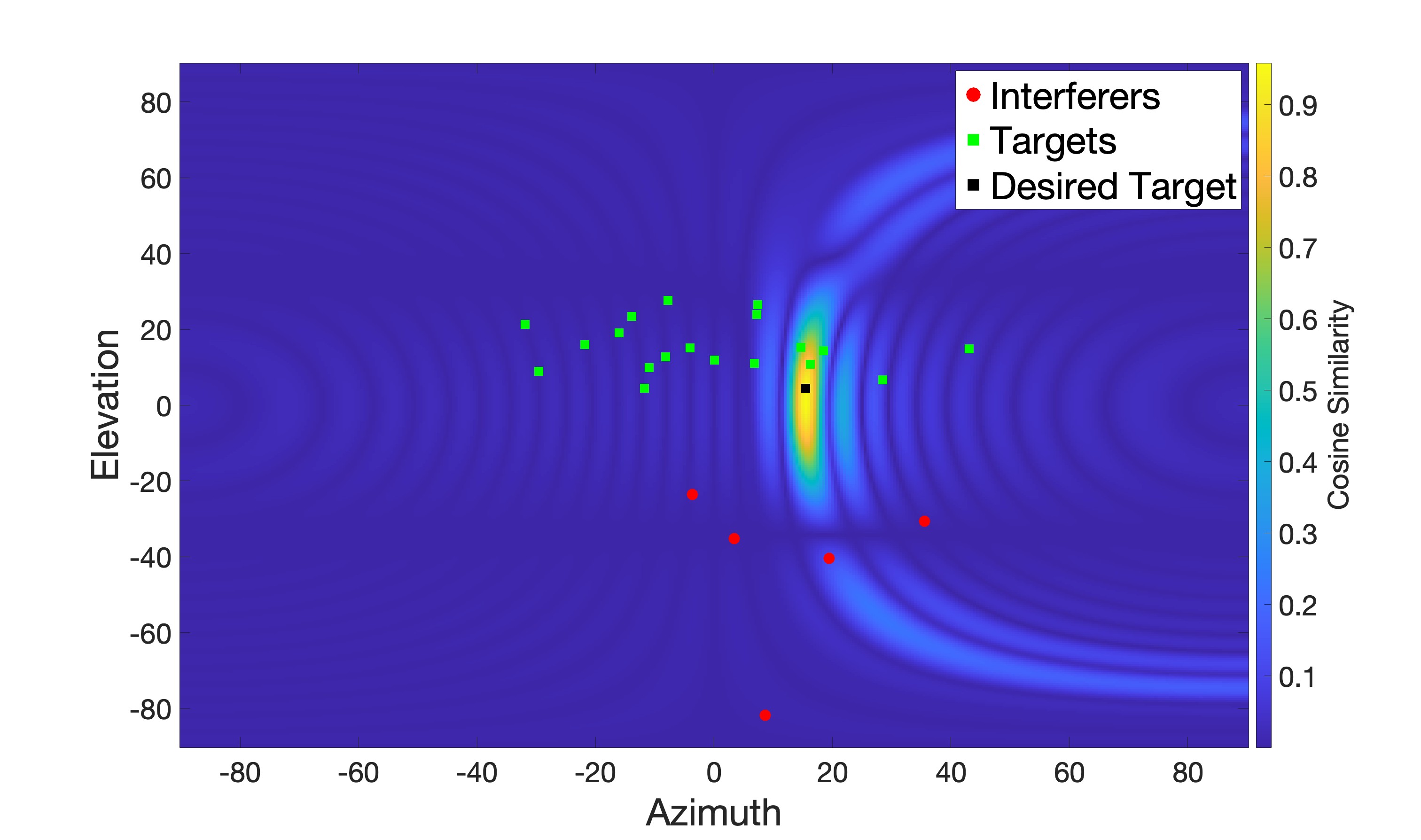}
        \caption{\footnotesize{Windowed Beamspace (\(W_z \times W_x = 2 \times 4\))}}
        \label{Fig:BP_A1_B}
    \end{subfigure}
    \caption{\small Beamforming pattern comparison for one of the targets in scenario \(\text{A}_1\).}
    \label{Fig:BP_A1}
\end{figure}

\begin{figure}[t]
    \centering
    % Second row of subfigures
    \begin{subfigure}{0.48\textwidth}
        \centering
        \includegraphics[width=\linewidth, height=0.6\linewidth]{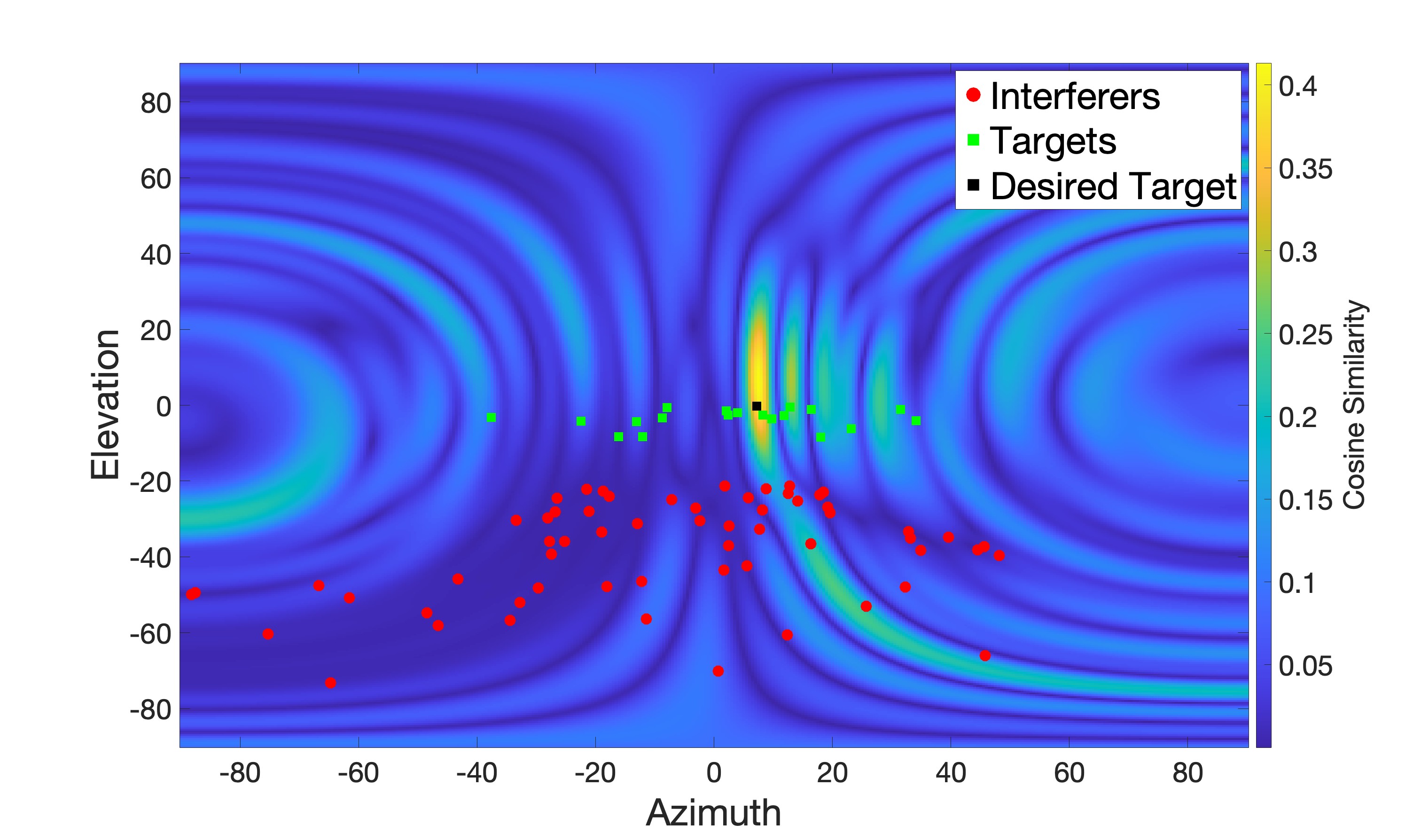}
        \caption{\footnotesize{Antenna space}}
        \label{Fig:BP_E2_A}
    \end{subfigure}
    \begin{subfigure}{0.48\textwidth}
        \centering
        \includegraphics[width=\linewidth, height=0.6\linewidth]{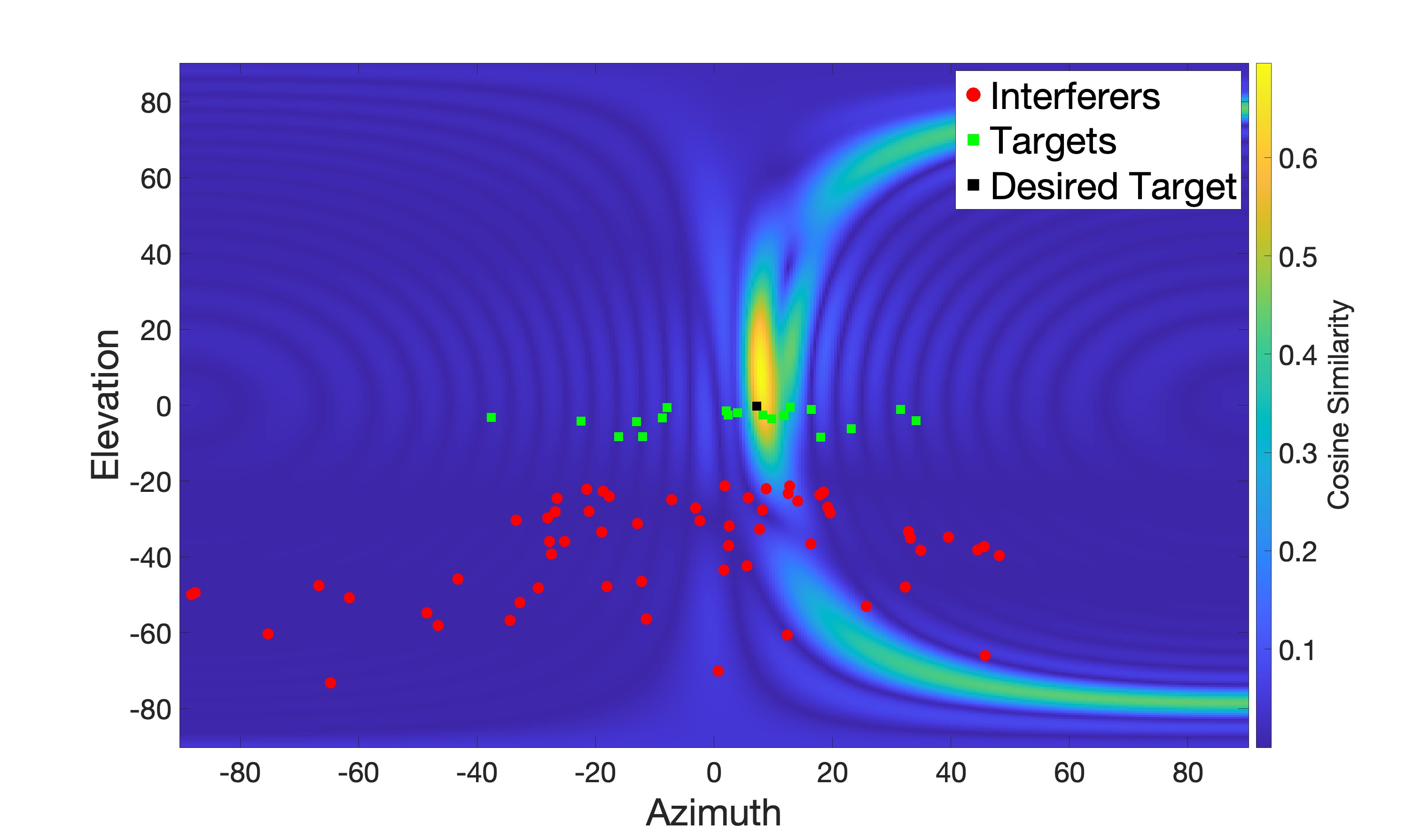}
        \caption{\footnotesize{Windowed Beamspace (\(W_z \times W_x = 4 \times 4\))}}
        \label{Fig:BP_E2_B}
    \end{subfigure}
    \caption{\small Beamforming pattern comparison for one of the targets in scenario \(\text{E}_2\).}
    \label{Fig:BP_E2}
\end{figure}

Figures~\ref{Fig:BP_A1} and~\ref{Fig:BP_E2} illustrate the beamforming patterns for selected targets in Scenarios \( \text{A}_1 \) and \( \text{E}_2 \), respectively. Since MVDR beamforming is performed independently for each subband, we visualize the beam patterns using the correlator corresponding to the center subband. These plots show that the windowed beamspace MVDR produces comparable or even improved beam patterns, particularly in terms of interference suppression. This improvement stems from the inherent concentration of energy in the beam domain, followed by spatial windowing, which enables deeper nulls—especially at angles farther from the main lobe. In low-interference scenarios like \( \text{A}_1 \), a small window size such as \( 2 \times 4 \) is sufficient to isolate the desired target and suppress interferers. However, in more challenging scenarios like \( \text{E}_2 \), where the beamspace is densely populated, the reduced dimensionality can result in performance degradation unless sufficient window size is allocated, especially along the vertical dimension.

\section{conclusion}
\label{Sec:conclusion}
We proposed a scalable windowed beamspace MVDR framework for wideband massive MIMO radar, aiming to reduce the computational and training complexity of adaptive processing without sacrificing detection performance. By transforming each subband signal into the beam domain and applying fixed low-dimensional spatial windows, our method achieves substantial dimension reduction. The results show that, with careful window selection, beamspace MVDR can match or even outperform full-dimensional MVDR in terms of detection accuracy—especially in low-interference scenarios—while maintaining significantly lower complexity.

We analyzed the impact of beamspace parameters such as FFT size and window dimensions, revealing their roles in managing interference leakage and ensuring robust performance under varying scene complexity. Experimental results on a challenging airborne dataset validate the efficacy of our approach across multiple operational regimes.
 
In future work, we plan to integrate the proposed beamspace framework into a modular architecture such as the tiled beamspace design in~\cite{han2024tiled}, enabling distributed processing with reduced complexity for scaling to even larger arrays. We also aim to investigate hardware-signal processing co-design for energy-efficient implementations balancing performance, computational complexity, and communication overhead.

\section*{Acknowledgment}
This work was supported in part by the Defense Advanced Research Projects Agency (DARPA) SOAP program, and in part by the Center for Ubiquitous Connectivity (CUbiC), sponsored by Semiconductor Research Corporation (SRC) and DARPA under the JUMP 2.0 program.

% \clearpage
\bibliographystyle{IEEEtran}
\bibliography{ref}

\end{document}